\documentclass[useAMS,usenatbib,fleqn]{mnras}

\usepackage{psfrag,amsmath,amssymb,color,epsfig,rotating}
\usepackage{graphicx}
\usepackage{graphics}
\usepackage{multirow}
\usepackage{tabularx}
\usepackage{geometry}
\usepackage{array}
\usepackage{comment}
\usepackage{setspace}

\usepackage[T1]{fontenc}
\usepackage{ae,aecompl}

\usepackage[dvipsnames]{xcolor}

\def\HI{{\rm HI}\,} 
\def\cth{{\cos \theta}\,} 
\def\mpci{\,{{\rm Mpc}^{-1}}}

\def\kk{\mathbf{k}} 
\def\kpe{k_{\perp}}
\def\kpa{k_{\parallel}}
\def\ka{{\bf k_1}}
\def\kb{{\bf k_2}}
\def\kc{{\bf k_3}}

\newcommand*{\id}{{\rm\hbox{1\kern-0.15em \vrule width .1pt depth-.2pt}}}

\title [Post-reionization 21-cm bispectrum]{Modelling the post-reionization neutral hydrogen ({\rm HI}) 21-cm bispectrum} 

\author[Sarkar et al.]{Debanjan
	Sarkar$^{1}$\thanks{debanjan@cts.iitkgp.ernet.in}, Suman Majumdar$^{3,4}$\thanks{suman.majumdar@iiti.ac.in}, Somnath Bharadwaj$^{1,
		2}$\thanks{somnath@phy.iitkgp.ernet.in}\\ $^1$Centre for Theoretical Studies, Indian
	Institute of Technology Kharagpur, Kharagpur - 721302, India \\ $^2$Department of Physics, Indian Institute of Technology Kharagpur,
	Kharagpur - 721302, India\\
	$^3$Discipline of Astronomy, Astrophysics and Space Engineering, Indian Institute of Technology Indore, Simrol, Indore 453552, India\\
	$^4$Department of Physics, Blackett Laboratory, Imperial College, London SW7 2AZ, U. K.\\}

\date{}
\pubyear{2019}

\begin{document}
	\label{firstpage}
	\pagerange{\pageref{firstpage}--\pageref{lastpage}}
	\maketitle

	\begin{abstract}
		Measurements of the post-reionization 21-cm bispectrum 
		$B_{\HI}(\mathbf{k_1},\mathbf{k_2},\mathbf{k_3})$  
		using various  upcoming  intensity mapping experiments hold the potential for 
		determining  the cosmological parameters at a high level of precision.  In this 
		paper we have estimated the 21-cm bispectrum in the $z$ range $1 \le z \le 6$  
		using semi-numerical simulations of the  neutral hydrogen ({\rm HI}) distribution.   
		We determine the $k$ and $z$ range where the 21-cm bispectrum can be adequately 
		modelled using the predictions of second order perturbation theory, and we use 
		this to predict the redshift evolution of the linear  and quadratic  {\rm HI} bias 
		parameters $b_1$ and $b_2$ respectively. The $b_1$ values are found to decreases 
		nearly linearly with decreasing $z$, and are  in good agreement with earlier 
		predictions obtained by modelling the   21-cm  power spectrum $P_{\HI}(k)$. 
		The $b_2$ values fall sharply with decreasing $z$, 	becomes zero at $z \sim 3$ 
		and attains a nearly constant value $b_2 \approx -  0.36$ at $z<2$. 
		We  provide polynomial fitting formulas for $b_1$ and $b_2$ as functions of $z$. 
		The modelling presented here is expected to be useful in future efforts to determine
		cosmological parameters and constrain primordial non-Gaussianity  using the 21-cm
		bispectrum.
	\end{abstract}
	

	\section{Introduction}
	\label{sec:introduction}

	The 21-cm radiation which originates from  the hyperfine transition in the ground
	state of the neutral hydrogen (\HI) atom offers a distinct way of mapping the
	large-scale structures (LSS)  over a large redshift range in the post-reionization 
	$(z <6)$ era. Here the collective 21-cm emission from the discrete individual \HI 
	sources appears as  a diffused background 	radiation below $1420$ MHz. 
	A statistical detection of the intensity fluctuations in this 21-cm background 
	is expected to quantify the underlying LSS \citep{bharadwaj-nath-sethi01}. 
	This technique is known as 21-cm intensity mapping, and this makes it possible 
	to survey large volumes of space using current and upcoming radio telescopes
	\citep{bharadwaj-sethi01,bharadwaj-pandey03,wyithe-loeb08-fluctuations-in-21cm}.
	In the post-reionization era, the absence of complex reionization
	processes make the 21-cm power spectrum proportional to the underlying 
	matter power spectrum \citep{wyithe-loeb09}. A detection of the 
	Baryon Acoustic Oscillation (BAO)  in the 21-cm power spectrum can be used 
	to place tight constraints on the dark energy equation of state
	\citep{chang-pen-peterson08, wyithe-loeb-geil08, masui-mcdonald-pen10, seo-dodelson10}.

	An accurate measurement of the 21-cm power spectrum also holds the possibility 
	of providing independent estimates of the different cosmological parameters
	\citep{loeb-wyithe08, bharadwaj-sethi-saini09, obuljen-castorina17}. 
	The cross-correlation of 21-cm signal with the other tracers of LSS like 
	the Lyman$-\alpha$ forest \citep{guhasarkar13-lya-21,carucci17-lya21,sarkar18-HI-lya}, 
	the Lyman-break galaxies \citep{navarro15-LBG21}, weak lensing 
	\citep{guhasarkar10-weaklens}, and  the integrated Sachs Wolfe effect 
	\citep{guhasarkar09-isw21}  have also been proposed  as 
	important cosmological probes in the post-reionization era.

	A statistical detection of the post-reionization \HI 21-cm signal was first
	reported in \citet{pen-21cmIM} through cross-correlation between the \HI Parkes All Sky Survey (HIPASS) and the six degree field galaxy redshift survey (6dfGRS). 
	In a subsequent work, \citet{chang-pen-bandura10} made a detection of the 21-cm 
	intensity mapping  signal in cross-correlation  between $z \approx 0.8$ 
	Green Bank Telescope (GBT)   observations and the DEEP2 optical galaxy redshift 
	survey. A similar detection of the cross-correlation signal between GBT observations
	and the WiggleZ Dark Energy Survey at $z \approx 0.8$	was reported by
	\citet{masui-switzer-banavar13}. \citet{switzer-masui-bandura-calin13}
	have measured the 21-cm auto-power spectrum using  GBT observations and they have
	constrained the amplitude of \HI fluctuations at $z \approx 0.8~$.

	Several low-frequency instruments like 
	the BAO from Integrated Neutral Gas Observations \citep[BINGO;][]{battye-brown12}, 
	the Canadian Hydrogen Intensity Mapping Experiment \citep[CHIME;][]{bandura14},
	the Tianlai Project \citep{chen-wang-16-tianlai}, 
	the HI Intensity Mapping program at the Green Bank Telescope 
	\citep[GBT-HIM;][]{chang16-GBT-HIM}, 
	the Five hundred metre Aperture Spherical Telescope \citep[FAST;][]{bigot16-FAST}, 
	the Australian Square Kilometre Array Pathfinder \citep[ASKAP;][]{johnston08-ASKAP}, 
	the Hydrogen Intensity and Real-time Analysis eXperiment 
	\citep[HIRAX;][]{kuhn19-HIRAX}
	are primarily planned to measure the BAO scale using the 21-cm signal in 
	the intermediate redshift range $z \sim 0.5-2.5~$. The linear radio-interferometric 
	array Ooty Wide Field Array \citep[OWFA;][]{subrahmanya-manoharan-chengalur17-OWFA} 
	is intended to measure the 21-cm power spectrum at $z \sim 3.35~$. On the other hand,
	instruments like the upgraded  Giant Metrewave Radio Telescope 
	\citep[uGMRT;][]{gupta17-uGMRT} and the Square Kilometre Array
	\citep[SKA;][]{santos-bull15-SKA-HI-IM-survey} have the potential to survey a large 
	redshift range in the post-reionization era.

	The 21-cm signal is intrinsically very weak (four to five orders of magnitude 
	smaller than the various astrophysical foregrounds), and it is really important to 	
	carefully model the signal in order to make accurate  predictions for the 
	detectability of the signal with the various telescopes ({\it e.g.}
	\citealt{bull-ferreira15-late-time-cosmology-with-21cm-IM, sarkar-bharadwaj-ali17-OWFA}). 
	Detailed modelling of the signal is also essential to interpret the detected 
	21-cm power spectrum and correctly estimate the cosmological model parameters
	\citep{obuljen-castorina17, hamsa-refregier19}. A precise modelling 
	of the 21-cm power spectrum is also required to understand the high-redshift 
	astrophysics \citep{IM-report2017}. Further, models for the expected 21-cm signal 
	are useful to validate  foreground removal and avoidance techniques
	({\it e.g.} \citealt{choudhuri17-thesis-FG} and references therein).

	A considerable amount of work has been carried out towards  modelling  the 
	the post-reionization \HI distribution and the expected 21-cm signal.
	Several efforts have been made to predict the 
	\HI power spectrum and bias using analytic
	prescriptions coupled with N-body simulations 
	\citep{bagla10, khandai-sethi-dimatteo11, tapomoy-mitra-majumder12}. 
	Subsequent works have also  used analytic prescriptions coupled 
	with smoothed particle hydrodynamics (SPH) simulations to study the \HI
	clustering \citep{villaescusa-navarro-viel-datta-choudhury14}. 
	Several cosmological hydrodynamic simulations have also been
	used to accurately model the galactic \HI content in the post-reionization universe \citep{dave-katz-openheimer13, barnes-haehnelt-14, villaescusa-navarro16-HI-in-galaxy-clusters}. 
	A number of analytical frameworks have also been developed to model the 21-cm 
	intensity mapping observables 
	\citep{hamsa-refregier17-halomodel1, hamsa-refregier-amara17-halomodel2, 
		castorina-villaescusa-navarro16, umeh-maartens-santosh16, umeh17, 
		aurelie-umeh-santos17-HIbias, hamsa-girish17}.

	In \citet{sarkar-bharadwaj-2016} (hereafter Paper I) we have used a semi-numerical technique, that combines dark-matter-only simulations and an analytic prescription
	to populate	the dark matter haloes with \HI \citep{bagla10}, to study the \HI clustering in the redshift range $1 \le z \le 6~$. We have quantified
	the \HI bias, through the signal power spectrum, across this $z$ range for $k$ values in the range 
	$0.04 \le k\,/{\rm Mpc}^{-1} \le 10~$, and we provide polynomial fitting formulas 
	for the bias across this $k$ and $z$ range.

	In \citet{sarkar-bharadwaj-2018}  and 
	\citet{sarkar-bharadwaj-2019} (hereafter Papers II and  III respectively), we have considered
	several  methods to incorporate the \HI peculiar velocities  in  
	semi-numerical simulations and use these to predict the 21-cm signal.  We  model
	the redshift-space 
	\HI power spectrum $P^s_{\HI}(\kpe,\kpa)$ with the assumption that 
	it can be expressed as a product of three terms: (i) the real space \HI 
	power spectrum $P_{\HI}(k) = b^2 (k)P(k)$, where $P(k)$ is the dark matter 
	power spectrum in real space and $b(k)$ is the \HI bias, (ii) a Kaiser
	enhancement \citep{kaiser87} factor and (iii) an independent Finger of God (FoG) 
	suppression \citep{jackson72-FoG} term. Considering a number of profiles for 
	the FoG suppression, we have found that the Lorentzian damping profile
	provides a reasonably good fit to the simulated $P^s_{\HI}(\kpe,\kpa)$ 
	across the entire $z$ range $(1 \le z \le 6)$ considered.

	In the simplest scenario, structure formation proceeds from Gaussian random initial conditions where the   different Fourier modes $\Delta(\kk)$ are uncorrelated and the statistics is completely specified by the power spectrum $P(k)$. It is however well known \citep{peebles80} that non-Gaussianity sets in as gravitational instability proceeds and the density fluctuations become non-linear.  The post-reionization \HI 21-cm signal is expected to be  significantly non-Gaussian at length-scales which have become non-linear. The phases of the different Fourier modes $\Delta(\kk)$ are now correlated, and in addition to the power spectrum it is now necessary to consider higher order statistics like the bispectrum $B(\ka,\kb,\kc)$  in order to quantify the statistics of the 21-cm signal.

	The study of the higher order statistics dates back to early measurements of 
	the galaxy three point correlation function using the Zwicky and the Lick  
	angular catalogues \citep{peebles76,groth77, fry82}. Subsequent studies of the 
	galaxy three point correlation function  include	
	\citet{bean83-3pt, efstathiou84-3pt, hale89-3pt, gaztanaga94-3pt}. More 
	recent studies \citep{cappi15,guo15-3pt,slepian17-boss, moresco17-vipers}
	have measured the galaxy three point correlation function at a high  level 
	of accuracy and these  have been  used to investigate the  galaxy morphology, 
	colour, and luminosity dependence.
	A number of theoretical frameworks 	have been developed to model the three point
	correlation and bispectrum using higher	order perturbation theory 
	\citep{fry84-bispec, fry94-3PCF, bharadwaj94,matarrese97, scoccimarro98-NL-bispec, 
		verde98-RS-bispec, scoccimarro00-bispec}. As proposed by \citet{matarrese97}, 
	the observed galaxy bispectrum \citep{feldman01-bispec, scoccimarro01-bispec, verde02-bispec,taruya07, gilmartin15-bispec} has been used to estimate the galaxy 
	bias parameters. The CMB bispectrum has been extensively studied, and recent 
	works \citep{planck15png,planck18png} have placed stringent bounds on the primordial
	non-Gaussianity.

	There have been several works  modelling the 21-cm bispectrum during various 
	stages of the cosmic evolution covering the dark ages
	\citep{pillepich07-dark-age}, cosmic dawn \citep{shimabukuro16-dawn}, 
	and epoch of reionization \citep{bharadwaj-pandey05, chongchitnan13, 
		yoshiura15, shimabukuro17, majumdar18-bispec, hoffman18}. In this paper we focus 
	on the post-reionization 21-cm signal which  is expected to be highly
	non-Gaussian due to the non-linear gravitational instability.  In an earlier paper 
	\citet{ali06-21cmbispec} have explored the 21-cm bispectrum signal expected at the GMRT.
	\citet{guhasarkar13-21cmbispec} have investigated the 	prospects of constraining 
	primordial non-Gaussianity  using measurements  of the 21-cm bispectrum, and also the 
	cross-correlation bispectrum of the 21-cm signal and  the Lyman$-\alpha$ forest. \citet{schmit19} have investigated  the possibility of measuring the 21-cm bispectrum 
	with several upcoming instruments. In addition to the bispectrum arising from gravitational
	instability, they have also considered the ISW-lensing contribution which they have found 
	to be orders of magnitude smaller. In a recent paper  
	\citet{ballardini19png} have investigated the prospects of constraining primordial non-Gaussianity using a combination of 21-cm observations, galaxy surveys and CMB lensing.

	Second order perturbation theory predicts \citep{matarrese97} that measurements of the
	bispectrum in the weakly non-linear regime can be used to determine the linear and 
	quadratic  bias parameters $b_1$ and $b_2$ respectively, and thereby break the degeneracy with the matter density parameter $\Omega_{m}$. It is anticipated that the upcoming
	post-reionization 21-cm intensity mapping experiments will be able  to measure both the 
	\HI 21-cm power spectrum and bispectrum  at a high level of sensitivity  leading to 
	precision measurements of the cosmological parameters including tight constraints on
	primordial non-Gaussianity. While the \HI 21-cm power spectrum has been extensively 
	studied and modelled using semi-numerical simulations, to the best of our knowledge 
	similar studies have not been carried out for the post-reionization 21-cm bispectrum.
	Such studies are necessary for precise quantitative predictions on the prospects of 
	detecting the 21-cm bispectrum using different upcoming instruments. In this paper 
	we have estimated the 21-cm bispectrum using semi-numerical simulations of the post-reionization \HI distribution. We determine the $k$ and $z$ range where the 
	21-cm bispectrum can be adequately modelled using the predictions of second order perturbation theory, and provide polynomial fitting formulas for $b_1$ and $b_2$ as 
	functions of $z$. The modelling presented here is expected to be useful in future efforts 
	to determine cosmological parameters using observations of the 21-cm intensity mapping 
	signal.

	The structure of this paper is as follows. In Section~\ref{sec:real-space-sim} we
	briefly describe the semi-numerical simulations of the  \HI distribution and discuss 
	the bispectrum estimator. In Section~\ref{sec:results} we present results for the 21-cm
	bispectrum estimated from the \HI simulations, and the modelling is presented in 
	Section~\ref{subsec:modelling}. Section~\ref{sec:discussion} contains the summary and discussion.

	Throughout this analysis we have used the best-fit cosmological parameters 
	from \cite{planck-collaboration15}.

	\section{Simulating the \HI 21-cm bispectrum}
	\label{sec:real-space-sim}
	
	We start by simulating the post reionization \HI distribution in real space.
	We have used a Particle Mesh {\it N}-body code \citep{bharadwaj-srikant04} to 
	generate the dark matter distribution in a comoving volume of 
	$[150.08\,{\rm Mpc}]^3$ in the redshift range $z = 1-6$ at a redshift interval 
	of $\Delta z = 0.5~$. The simulations have a spatial
	resolution of (comoving) $0.07\,{\rm Mpc}$ which roughly translates into a 
	mass resolution of $10^8\,\text{M}_{\odot}$. The simulations used here are the 
	same as those analysed in Paper I and Paper II. We have used the Friends-of-Friends
	(FoF) algorithm \citep{davis85} with a linking length of $0.2$ in units of 
	the mean interparticle separation to identify the assembly of dark matter 
	particles that form a halo.	We have assumed that a halo must contain a 
	minimum of $10$ dark matter particles which limits
	the halo mass resolution to $10^9\,\text{M}_{\odot}$. This halo mass resolution 
	is sufficient for the reliable prediction of the 21-cm brightness temperature
	fluctuation \citep{kim-wyithe-baugh-lagos16}.

	We then populate the dark matter haloes using an analytical prescription, 
	suggested by \citet{bagla10}, which considers that \HI in the post-reionization era 
	resides	solely inside the dark matter haloes. 
	\citet{bagla10} have 
	provided a redshift-dependent relation that connects the virialized halo mass 
	$M_h$ with the circular velocity $v_{\text{circ}}$ of the halo
	\begin{equation}
	M_h= 10^{10}\,\text{M}_{\odot}\, \left( \frac{v_{\text{circ}}}{60\,\text{km}\,\text{s}^{-1}}\right)^3
	\left( \frac{1+z}{4} \right)^{-\frac{3}{2}}\,.
	\label{eq:virial-relation}
	\end{equation}
	The prescription assumes that a halo will not be able to host \HI if its mass 
	is below a minimum threshold $M_{\text{min}}$. It is expected that these smaller 
	haloes would not 
	be able to shield the neutral gas from ionizing background radiation. The prescription
	further assumes that the \HI fraction of a halo will go down if its mass exceeds
	an upper limit $M_{\text{max}}$. This is based on the observations in the local 
	universe where the massive elliptical galaxies and the galaxy clusters
	contain very little \HI \citep{serra-oosterloo-morganti12}.
	The threshold values $M_{\text{min}}$ and $M_{\text{max}}$ at any redshift can be 
	calculated by setting $v_{\text{circ}}=30\, {\rm km\,s}^{-1}$ and 
	$v_{\text{circ}}=200\, {\rm km\,s}^{-1}$ respectively in
	Equation~\ref{eq:virial-relation}. According to the \HI
	assignment prescription, the \HI mass $M_{\HI}$ of a halo is related to the
	halo mass $M_h$ as 
	\begin{equation}
	M_{\HI}(M_h) = \begin{cases} f_3 \frac{M_h}{1+\left( \frac{M_h}{M_{\text{max}}} \right)}
	, & \mbox{if } M_h \geq M_{\text{min}} \\ 0, & \mbox{otherwise}\end{cases} \,,
	\label{eq:HI-prescription}
	\end{equation}
	where $f_3$ is a free parameter which decides the \HI content in our simulations.
	The choice of $f_3$ does not influence the results of this work, and have 
	used $f_3$ such that the cosmological \HI density parameter $\Omega_{\HI}$
	stays fixed at a value $\sim 10^{-3}$ in our simulations. Here we have placed all
	the \HI at the halo centre of mass. For a detailed discussion of the above 
	technique, the reader is referred to Paper I and Paper II.

	
	\begin{figure}
		\centering
		\includegraphics[width=0.35\textwidth,angle=0]{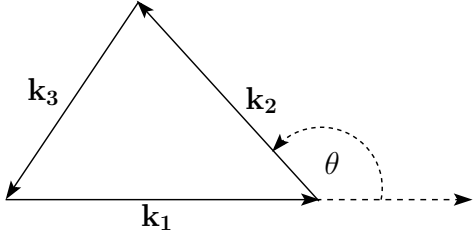}
		\caption{This shows a generalised closed triangle configuration in $\kk$ space
			that we have used for bispectrum estimation. Here $\theta$ is the angle
			between $\ka$ and $\kb$, and is defined as $\cos \theta=( \ka.\kb) /(k_1 k_2)$.}
		\label{fig:triangle}
	\end{figure}

	The steps described till now generates the real space \HI distribution. 
	We use the Cloud-in-cell (CIC) interpolation to calculate the $\HI$ density 
	contrast $\delta_{\HI}(\mathbf{x})$ on the grids. We use 
	$\Delta_{\HI}(\mathbf{k})$ which is the 
	Fourier transform of $\delta_{\HI}(\mathbf{x})$  to compute the \HI bispectrum.

	We define the bispectrum $B_{\HI}(\ka,\kb,\kc)$ of the post-reionization 
	\HI 21-cm signal through, 
	\begin{equation}
	\langle \Delta_{\HI}(\ka) \Delta_{\HI}(\kb) \Delta_{\HI}(\kc) \rangle = V \delta^{\text{K}}_{\ka+\kb+\kc,0} B_{\HI}(\ka,\kb,\kc)\,,
	\label{eq:definition}
	\end{equation}
	where, $\delta^{\text{K}}_{\ka+\kb+\kc,0}$ is the Kronecker delta function 
	which  is $1$ when $\ka+\kb+\kc=0$ and $0$ otherwise. This ensures that 
	only triplets $(\ka,\kb,\kc)$	which form a closed triangle 
	(see Figure~\ref{fig:triangle})  contribute to the bispectrum. Here $V$ is
	the simulation  volume (comoving). Note that, $B_{\HI}(\ka,\kb,\kc)$ here only 
	depends  on the	shape  and size of the triangle  and does not depend on how 
	the triangle is  oriented. Considering a triangle as shown in 
	Figure~\ref{fig:triangle}, we use the three parameters  $(k_1,n,\cth)$ to 
	uniquely specify its  shape and size. The three parameters are defined as 
	$k_1=\mid \ka \mid$,  
	\begin{equation}
	n=\frac{k_2}{k_1}\,,
	\label{eq:cons1}
	\end{equation}
	and 
	\begin{equation}
	\cth=\frac{\ka.\kb}{k_1k_2}\,
	\label{eq:cons2}
	\end{equation}
	where $-1 \leqslant \cth \leqslant 1$.
	We have used the binned bispectrum estimator presented in 
	\citet{majumdar18-bispec} (and also in \citealt{watkinson17-bispec}) to 
	calculate $B_{\HI}(k_1,n,\cth)$. The entire $k_1$ range
	of our simulation was divided into $15$ equal logarithmic bins,
	and the bins used here are exactly the same as those in 
	\citet{majumdar18-bispec}. In order to limit the length of the analysis,  	
	we have restricted the present study to three specific $n$ values $1,2$ and $5$.  
	Here $n=1$ corresponds to isosceles triangles for which $\cth=-0.5$	corresponds 
	to  an equilateral triangle.  For all values of $n$, 
	the extreme limits  $\cth \rightarrow -1$ and $\cth \rightarrow 1$  
	respectively  correspond to the squeezed ($k_3=  \mid 1-n \mid \, k_1$ ) 
	and the extended  ($k_3= (1+n) k_1$) triangles where in both cases the three 
	vectors $\ka,\kb$ and $\kc$ are colinear.

	We have generated five statistically independent realizations of the simulation 
	to estimate the mean and variance for all the results presented here.
	For convenience, we have considered the dimensionless form of the \HI bispectrum 
	\begin{equation}
	\Delta^3_{\HI}(k_1,n,\cth)=k_1^6 \,  B_{\HI}(k_1,n,\cth)\, /(2 \pi)^2 \,.
	\label{eq:dimless}
	\end{equation}

	\section{Results}
	\label{sec:results}
	
	\begin{figure*}
		\centering
		\includegraphics[width=0.98\textwidth,angle=0]{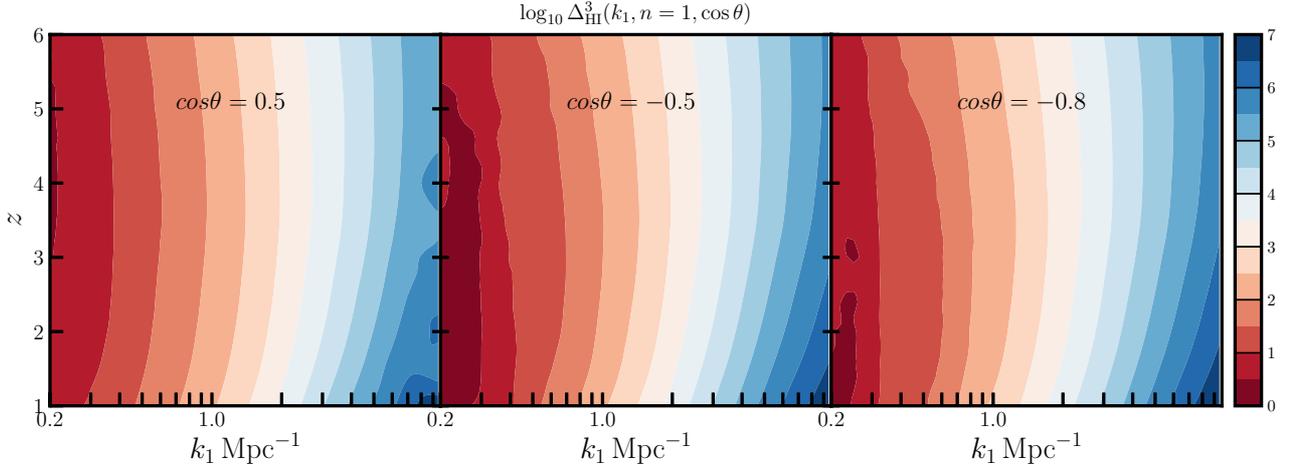}
		\caption{This shows the joint $k_1$ and $z$ dependence of 
			$\Delta^3_{\HI}(k_1,n,\cth)$ for the isosceles triangles 
			({\it i.e.} $n=1$) at three different values of $\cth$.}
		\label{fig:bispec_k1_z}
	\end{figure*}

	Figure~\ref{fig:bispec_k1_z} shows the joint $k_1$ and $z$ dependence of 
	$\Delta^3_{\HI}(k_1,n,\cth)$ for  isosceles triangles ({\it i.e.} $n=1$)  
	at three different values of $\cth$. We first discuss the central panel which 
	considers $\cth=-0.5$ that corresponds to equilateral triangle. 
	The equilateral triangle is a special case where the three Fourier modes that 
	appear in the bispectrum (Equation~\ref{eq:definition}) have equal magnitude 
	$(k_1=k_2=k_3)$.  Considering  any  fixed redshift $z$, we see that 
	$\Delta^3_{\HI}(k_1,n,\cth)$ increases monotonically with 
	increasing $k_1$. We notice  that the $\Delta^3_{\HI}(k_1,n,\cth)$
	contours are nearly vertical in the range  $k_1<0.4 \mpci$ 
	which implies $\Delta^3_{\HI}(k_1,n,\cth)$ does not vary 
	much with redshift in this $k$ range. However, the $\Delta^3_{\HI}(k_1,n,\cth)$ 
	contours are inclined and also curved for $k_1>0.4 \mpci$ indicating a significant 
	variation with $z$.   Considering a fixed  value of $k_1$ in the range 
	$k_1>0.4 \mpci$ we see that,  $\Delta^3_{\HI}(k_1,n,\cth)$ first decreases 
	with increasing $z$, then  becomes minimum at $z\sim 3.5$ and increases again 
	at $z>4$. The left-hand and right-hand panels  show results for $\cth=0.5$ and $\cth=-0.8$ 
	which corresponds to obtuse and acute triangles respectively. 
	We see that for both the obtuse and acute triangles the behaviour of  
	$\Delta^3_{\HI}(k_1,n,\cth)$ is similar  to that of the equilateral triangle. 
	These results are also very similar if we consider  $n=2$ and $n=5$ and we have 
	not shown these here.

	\begin{figure*}
		\centering
		\includegraphics[width=0.98\textwidth,angle=0]{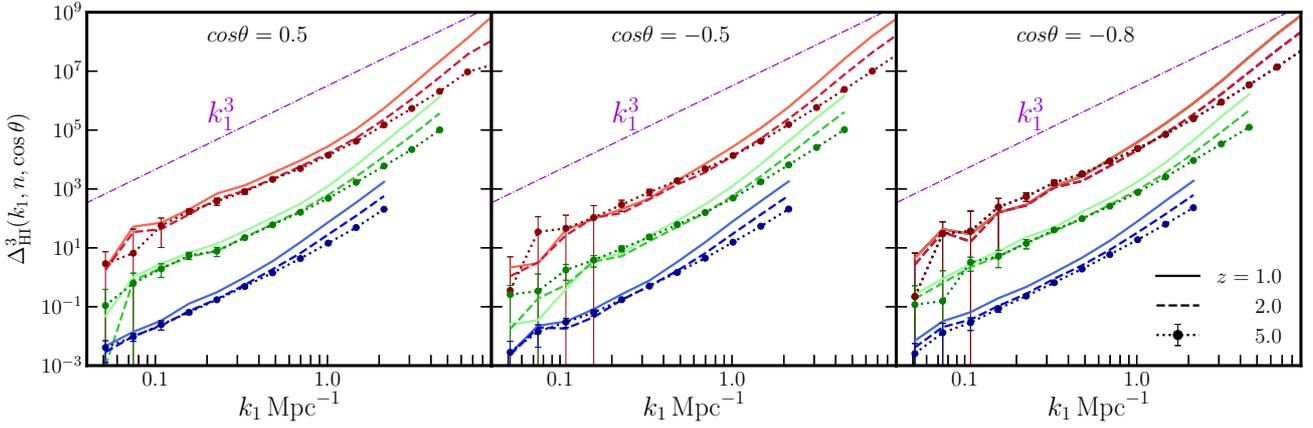}
		\caption{This shows $\Delta^3_{\HI}(k_1,n,\cth)$ as a function of $k_1$ for 
			the same three values of $\cth$ ($0.5,-0.5,-0.8$) as those considered in 
			Figure~\ref{fig:bispec_k1_z}. In each panel, the results 
			for the $n=1,2$ and $5$ triangles are plotted together. The results for the 
			different $n$ values overlap for almost the entire $k_1$ range, and we have 
			multiplied the results for $n=1$ and $2$ with $100$ and $10$ respectively 
			in order to show them clearly. We also show the power law $\sim k_1^3$ for
			convenience.
		}
		\label{fig:k1_dependence}
	\end{figure*}

	The three panels of Figure~\ref{fig:k1_dependence} show $\Delta^3_{\HI}(k_1,n,\cth)$
	as a function of $k_1$ for the same three values of $\cth$ ($0.5,-0.5,-0.8$) as 
	those considered in Figure~\ref{fig:bispec_k1_z}. In each panel, the results 
	for the $n=1,2$ and $5$ triangles are plotted together. The results for the 
	different $n$ values overlap for almost the entire $k_1$ range, and we have 
	multiplied the results for $n=1$ and $2$ with $100$ and $10$ respectively 
	in order to show them clearly. In all cases 
	the amplitude and $k_1$ dependence of $\Delta^3_{\HI}(k_1,n,\cth)$ 	 
	is almost constant with   redshift at $z \ge 3$, and consequently we show 
	the results at only three  redshifts $z=1,2$ and $5$.   Considering all 
	the results together, we see that $\Delta^3_{\HI}(k_1,n,\cth)$ exhibits 
	a power law behaviour 
	$\Delta^3_{\HI} \sim k_1^{\alpha}$ for $k_1<1 \mpci$
	with $\alpha \approx 3$ which is nearly independent of redshift, only 
	the amplitude of $\Delta^3_{\HI}(k_1,n,\cth)$ changes with redshift. 
	However this power law does	not hold at small $k_1$  ($<0.1 \mpci$) where 
	the cosmic variance dominates.  We also see that  $\Delta^3_{\HI}(k_1,n,\cth)$ 
	is steeper than $k_1^{3}$ at large $k_1$ ($>1 \mpci$) where the non-linear	
	effects are important, and this steepening increases at lower redshifts.  
	We also note that for $n=5$ the steeping sets in at a smaller value of $k_1$ 
	as compared to $n=1$. This is consistent with the notion that the steepening 
	arises from the small-scale non-linear effects which are expected to be more 
	pronounced for $n=5$ ($k_2=5 k_1$) as compared to $n=1$.

	\begin{figure*}
		\centering
		\includegraphics[width=0.98\textwidth,angle=0]{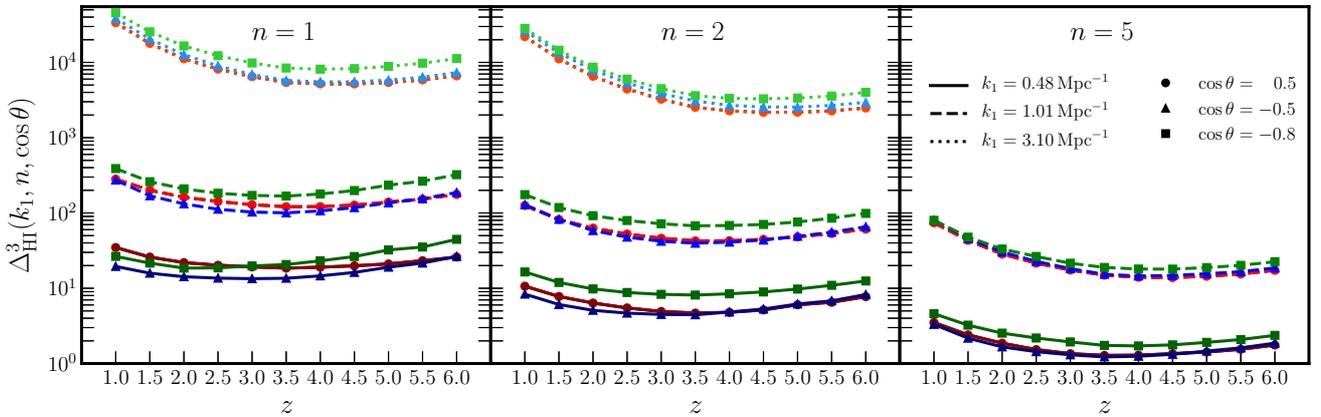}
		\caption{This shows the variation of $\Delta^3_{\HI}(k_1,n,\cth)$
			with redshift  at fixed values of $k_1$ and $\cth$.  We consider three
			representative $k_1$ values $(0.48,1.01,3.10) \mpci $, and  $\cth=(0.5,-0.5,-0.8)$ 
			same as in the earlier figures. The left-hand, central and right-hand 
			panels show results 
			for $n=1,2$ and $5~$ respectively. Note that  for $n=5$  we do not have results at 
			$k_1 = 3.10\mpci$  as the corresponding $k_2(=5k_1)$ is beyond the range of our 
			simulations.}
		\label{fig:z_variation}
	\end{figure*}

	Figure~\ref{fig:z_variation} shows the variation of $\Delta^3_{\HI}(k_1,n,\cth)$
	with redshift  at fixed values of $k_1$ and $\cth$.  We consider three
	representative $k_1$ values $(0.48,1.01,3.10) \mpci $ which correspond to 
	progressively increasing non-linear effects, and  $\cth=(0.5,-0.5,-0.8)$ 
	same as in the earlier figures. 
	The left-hand, central and right-hand panels show results for $n=1,2$ and $5~$ respectively.
	Note that  for $n=5$  we do not have results at $k_1 = 3.10\mpci$  as the 
	corresponding $k_2(=5k_1)$ is beyond the range of our simulations. 
	We see that the value of $\Delta^3_{\HI}(k_1,n,\cth)$ increases significantly 
	if $k_1$ increases, decreases to some extent if $n$ is increased, and  
	shows a relatively smaller variation with $\cth$. 
	Considering the overall  $z$ dependence,  we see that with  decreasing 
	$z$  the value of $\Delta^3_{\HI}(k_1,n,\cth)$ initially decreases,  shows
	a minima at $z \sim 3.5$ and then increases at lower redshifts. This redshift 
	evolution is relatively weak at the smallest $k_1$ which is weakly non-linear 
	at $z \le2$ (see Figure 2 of Paper I). The redshift evolution   is most pronounced 
	at the largest $k_1$  which is strongly non-linear at $z\le 3$ 	where
	$\Delta^3_{\HI}(k_1,n,\cth)$ shows  a noticeable increase with decreasing $z$.  	
	We also find a more pronounced redshift evolution for the larger  $n$ where the
	non-linear effects are expected to be stronger.   We note that the bispectrum of 
	the underlying matter distribution is expected to increase monotonically as $z$
	decreases. In contrast, we see that the \HI bispectrum first declines and then 
	increases as $z$ decreases. This, as we shall see later, can be understood in 
	terms of the redshift evolution of the \HI bias.

	\begin{figure*}
		\centering
		\includegraphics[width=0.98\textwidth,angle=0]{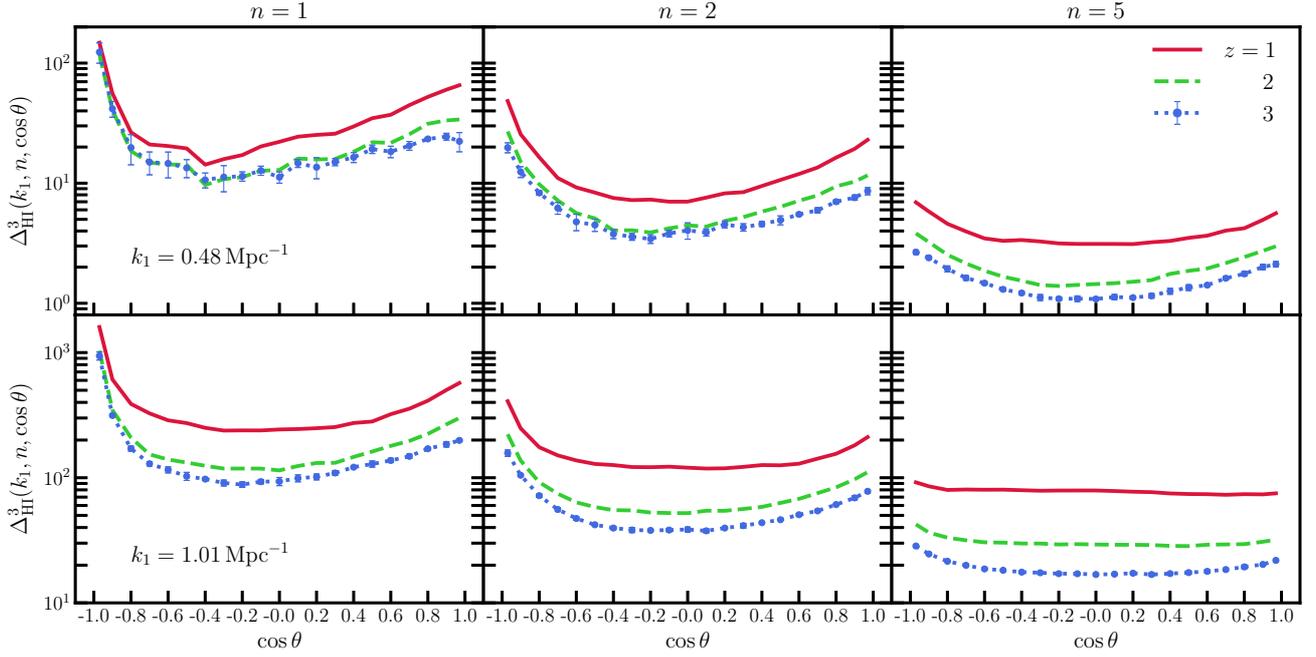}
		\caption{This shows $\Delta^3_{\HI}(k_1,n,\cth)$ as a function of $\cth$ at three 
			different redshifts $z=1,2$ and $3$. The top and bottom rows   
			show results at $k_1=0.48$ and $1.01\mpci$ respectively, while the  
			left, central  and right columns  show results  for $n=1,2$ and $5$ 
			respectively. The $1-\sigma$ spread estimated from five statistically
			independent realizations of the simulations is shown for results
			at $z=3$.}
		\label{fig:shape_dependence}
	\end{figure*}

	Figure~\ref{fig:shape_dependence} shows $\Delta^3_{\HI}(k_1,n,\cth)$ as a 
	function of $\cth$ at three different redshifts $z=1,2$ and $3$, we have 
	similar  results at higher redshifts where  the curves overlap and we 
	have not shown these here. The top and bottom rows   
	show results at $k_1=0.48$ and $1.01\mpci$ respectively, while the  
	left, central  and right columns  show results  for $n=1,2$ and $5$ respectively.
	Considering all the panels together, in nearly all cases we have a ``U" shaped 
	curve relating     $\Delta^3_{\HI}(k_1,n,\cth)$ to $\cth$. The minima appears 
	to be independent of $k_1$ and $z$, and shifts from  
	$\cth \approx -0.5$ to $-0.2$ and $0$ for $n=1,2$ and $5$. 
	The ``U" shaped $\cth$ dependence is well pronounced for $n=1$ and $2$ 
	at the higher redshifts where the $\kk$ modes involved are weakly non-linear. 
	However,  the ``U"  is rather broad and sometimes nearly flat in many cases, 
	particularly at low $z$ $(=1)$ or  if either of $k_1$ or $n$ is increased and 
	the $\kk$ modes involved are strongly non-linear.  It is difficult to reliably 
	estimate the minima in these cases.

	\section{Modelling the \HI Bispectrum}
	\label{subsec:modelling}
	
	We assume that the \HI traces the dark matter with a possible bias. Here  we retain terms
	upto the quadratic order in the relation between \HI density contrast  
	$\delta_{\HI}$ and the dark matter density contrast $\delta$
	\begin{equation}
	\delta_{\HI}=b_1 \delta + \frac{b_2}{2} \delta^2\,,
	\label{eq:NL-biasing}
	\end{equation}
	where $b_1$ and $b_2$ are the linear and quadratic bias parameters respectively.
	The linear bias $b_1$ can be calculated using  the
	relation
	\begin{equation}
	P_{\HI}= b_1^2 P\,,
	\label{eq:defn-b1}
	\end{equation}
	where $P_{\HI}$ and $P$ are respectively the \HI and the dark matter power 
	spectra. Paper I  presents a detailed analysis of the linear \HI bias allowing 
	for the possibility that $\tilde{b}_1(k)$,  defined in Fourier space, is both 
	complex and $k$ dependent.  The analysis however shows that it is adequate 
	to assume a real scale-independent bias at small $k$.  
	The quadratic bias  $b_2$ introduces non-linearity in the relation between
	$\delta_{\HI}$ and $\delta$. Here $b_1$ and $b_2$  both have been assumed  
	to be real scale-independent quantities which  vary only with redshift.

	Considering second order \citep{fry84-bispec} perturbation theory, 
	\citet{matarrese97} and \citet{ scoccimarro00-bispec} have calculated 
	the bispectrum for a biased tracer of the underlying dark matter 
	distribution. Applying this to the \HI field  (Equation~\ref{eq:NL-biasing}) 
	we have 
	\begin{multline}
	B_{\HI}(\ka,\kb,\kc)=b_1^3 \left[ 2F(\ka,\kb) P(k_1) P(k_2) + \text{cyc.} \right]
	\\+ b_1^2 b_2 \left[ P(k_1) P(k_2) + \text{cyc.} \right]\,,
	\label{eq:HI-bispec-th}
	\end{multline}
	where 
	\begin{multline}
	F(\ka,\kb)= \left( \frac{1+\kappa}{2}\right) + \left( \frac{\ka.\kb} {2 k_1 k_2} \right)
	\left(  \frac{k_1}{k_2} + \frac{k_2}{k_1} \right) \\ +  \left( \frac{1-\kappa}{2}\right)
	\left( \frac{\ka.\kb}{ k_1 k_2} \right)^2\,,
	\label{eq:F-kernel}
	\end{multline}
	with  $\kappa=\frac{3}{7} \Omega_m^{-\frac{1}{143}}$ for the
	$\Lambda CDM$ cosmology. Here we find that it is convenient to model 
	the \HI bispectrum $B_{\HI}(\ka,\kb,\kc)$ in terms of the \HI power spectrum 
	$P_{\HI}(k)$ using Equation~\ref{eq:defn-b1} whereby we obtain 
	\begin{multline}
	B_{\HI}(\ka,\kb,\kc)= \frac{1}{b_1} \left[ 2F(\ka,\kb) P_{\HI}(k_1) P_{\HI}(k_2) + \text{cyc.} \right]
	\\+ \frac{b_2}{b_1^2} \left[ P_{\HI}(k_1) P_{\HI}(k_2) + \text{cyc.} \right]\,.
	\label{eq:HI-bispec-th2}
	\end{multline}
	We have used Equation~\ref{eq:HI-bispec-th2} to model the \HI bispectrum $B_{\HI}(\ka,\kb,\kc)$.  The \HI power spectrum $P_{\HI}(k)$ is known from 
	simulations and this has been studied in Paper I. The model therefore has only 
	two free parameters, namely the two \HI  bias parameters  $b_1$ and $b_2$. 
	The second  term in the r.h.s. of Equation~\ref{eq:HI-bispec-th2}  depends 
	only on the magnitude of the three modes $\ka,\kb$ and $\kc$ whereas the first 
	term also depends on the shape of the triangle through $F(\ka,\kb)$. 
	This feature allows us to separately determine $b_1$ and $b_2$ by fitting the 
	model (Equation~\ref{eq:HI-bispec-th2}) to the \HI bispectrum obtained from the
	simulations.  We note that the model used here  is entirely based on second order
	perturbation theory which is expected to hold  only in the weakly non-linear regime.
	Many of the modes considered here are however in the strongly non-linear regime 
	where we do not expect the model to provide an adequate description of the \HI
	bispectrum.  We therefore first need to establish the range of Fourier modes
	$(\ka,\kb,\kc)$ where the model (Equation~\ref{eq:HI-bispec-th2}) provides an 
	adequate description of the \HI bispectrum obtained from the simulations.

	\begin{table}
		\centering
		{\renewcommand{\arraystretch}{1.1}%
			\begin{tabular}{c c c}
				\hline
				$z$	& $n$ & $k_1 \mpci$ \\ \hline
				\multirow{3}{*}{$1$} & $1$ & $0.33,0.48$ \\  
				& $2$ & $0.33$ \\ 
				& $5$ & $0.16$ \vspace{0.3cm} \\
				
				\multirow{3}{*}{$1.5-2$} & $1$ & $0.33,0.48$ \\  
				& $2$ & $0.33,0.48$ \\  
				& $5$ & $0.16$ \vspace{0.3cm} \\ 
				
				\multirow{3}{*}{$2.5-4.5$} & $1$ & $0.33,0.48,0.70$ \\  
				& $2$ & $0.33,0.48,0.70$ \\  
				& $5$ & $0.16$ \vspace{0.3cm} \\
				
				\multirow{3}{*}{$5.0 - 6.0$} & $1$ & $0.33,0.48,0.70$ \\ 
				& $2$ & $0.33$ \\  
				& $5$ & $0.16$ \\ 
				\hline
		\end{tabular} }
		\label{tab:k1-range}
		\caption{This presents the $k_1$ and $n$ values	that we have used for 
			fitting the \HI bispectrum  at different redshifts.}
	\end{table}

	\begin{figure*}
		\centering
		\includegraphics[width=0.98\textwidth,angle=0]{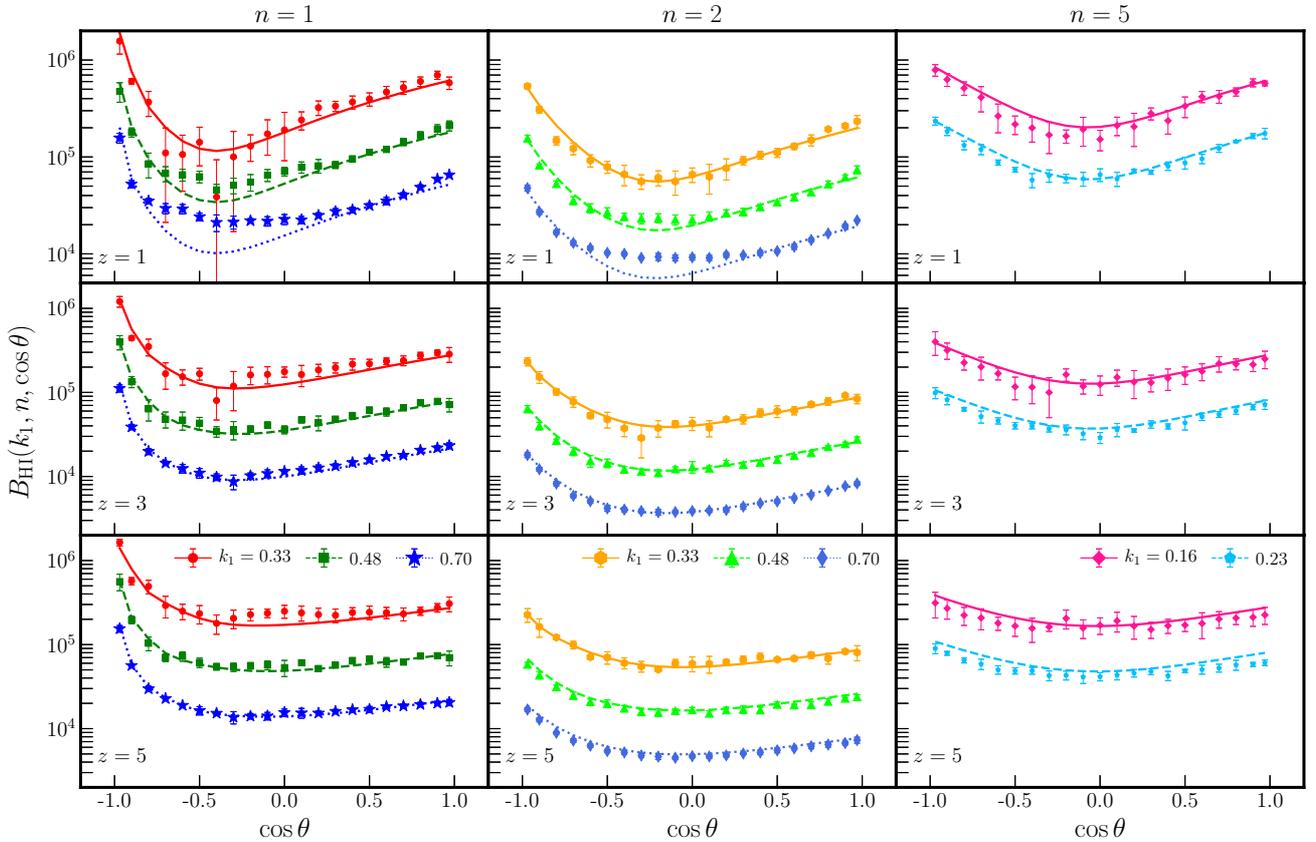}
		\caption{This shows the best fit prediction of our model
			(Equation~\ref{eq:HI-bispec-th2}) for the \HI bispectrum
			along with the simulated values  for various  triangle configurations.
			The top, middle and bottom rows show results at $z=1,3$ and $5$, while
			the left, central and right columns show results for $n=1,2$ and $5$.
			For each $n$ value, results for different $k_1$ (in unit of $\mpci$)
			are shown with different symbols and line styles. 
		}
		\label{fig:best-fit-bispec}
	\end{figure*}

	We have performed a  $\chi^2$ minimization with respect to $b_1$ and $b_2$ 
	in order to determine the parameter values for which the model (Equation~\ref{eq:HI-bispec-th2}) best  fits 
	$B_{\HI}(k_1,n,\cth)$ obtained from the simulations. Note that we 
	have used $P_{\HI}(k)$ obtained in Paper I for the fits. As mentioned earlier, 
	the bias parameters $b_1$ and $b_2$  are assumed to evolve with redshift and 
	we have separately carried out the fitting at each redshift. 
	We have initially attempted to fit $B_{\HI}(k_1,n,\cth)$ using the 
	entire $(k_1,n,\cth)$ range available in the simulations,  we however 
	find that the best fit reduced $\chi^2$ is rather large indicating a poor fit.
	While our model is expected to perform well at small $k$
	(large scales), we find that the sample variance is quite large for the first 
	few $k_1$ bins and hence we do not use these  for the subsequent fitting. 
	We also do not expect our	model to work at large  $k$ which are strongly 
	non-linear, and we find  that our model	predictions deviate significantly 
	from the simulated values. At each redshift, for   the various combinations 
	of  $k_1$ and $n$ we have individually considered $B_{\HI}(k_1,n,\cth)$  as 
	a function of $\cth$, some of these are shown in  Figure~\ref{fig:best-fit-bispec}. 
	We find a good fit with reduced $\chi^2$ of the order of unity provided we restrict 
	the $k_1, \, n$ values to the range shown in Table~\ref{tab:k1-range}, we however
	consider the full range $-1 < \cth < 1$  throughout. Note that the $k_1, \, n$ 
	range varies with $z$. The non-linear effects increase at lower redshifts, and 
	the combinations with larger $k_1$ and $n$ values are progressively dropped at 
	lower redshifts.

	\begin{figure}
		\centering
		\includegraphics[width=0.47\textwidth,angle=0]{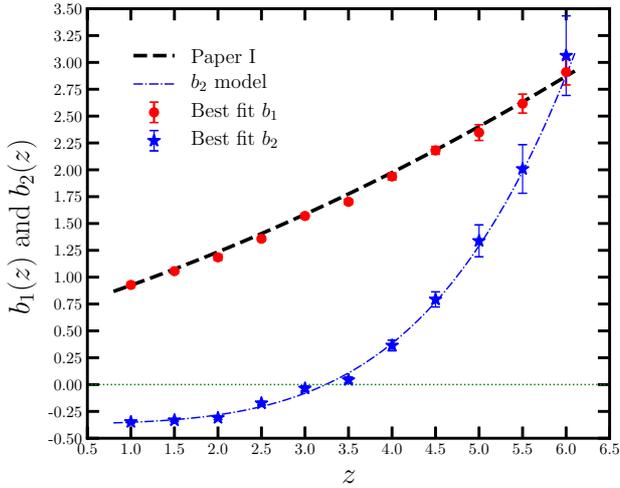}
		\caption{This shows the best-fitting values of the bias parameters  $b_1$ 
			(red circles) and
			$b_2$ (blue stars) at different redshifts. Error bars show the $1-\sigma$ 
			fitting uncertainties for the respective quantities. The black dashed line shows $b_1$ calculated using Equation~\ref{eq:linear_bias_model} whereas the blue line shows the  prediction of Equation~\ref{eq:model_b2} for $b_2$.}
		\label{fig:best-fit-biases}
	\end{figure}

	Figure~\ref{fig:best-fit-bispec} shows the best fit prediction of our model
	(Equation~\ref{eq:HI-bispec-th2}) for the \HI bispectrum
	along with the simulated values  for various  triangle configurations. 
	Considering the $n=1$ and $2$  triangles, we see that the model provides 
	a reasonably good fit to all the simulated values 
	shown here  except $k_1=0.70 \mpci$  at  $z=1$  where  the model  
	underpredicts the bispectrum.   Considering the $n=5$ triangles, 
	we see that the model provides a reasonably good fit to the simulations  
	for $k_1=0.16 \mpci$ at all redshifts, the model overpredicts the 
	bispectrum at other $k_1$ values.

	Figure~\ref{fig:best-fit-biases} shows the best-fitting values of the two 
	bias parameters  $b_1$ and $b_2$ at different redshifts. We see that 
	the best fit linear \HI bias $b_1$ falls almost linearly with decreasing
	$z$. In Paper I we have determined $b_1$ using the ratio of the power spectra 
	(Equation~\ref{eq:defn-b1}) allowing for both the scale  $(k)$ dependence and 
	redshift evolution, and we have modelled the joint $k$ and $z$ dependence 
	of $b_1(k,z)$  using a polynomial	which is quartic in $k$ and quadratic 
	in $z$ (Equation A1 of Paper I). In the present work  
	we  consider the  $k \rightarrow 0$ limit of $b_1(k,z)$ obtained in Paper I 
	to predict the scale-independent component of $b_1$ (see Figures 4 and 5 of Paper I) 
	whose  $z$ dependence can be modelled  using the polynomial 
	\begin{equation}
	b_1(z)=b_{10}+b_{11}z+b_{12}z^2\,,
	\label{eq:linear_bias_model}
	\end{equation}
	where the  coefficients have values  $(b_{10},b_{11},b_{12})=(0.653,0.252,0.0196)$. 
	We have	plotted the predictions of Equation~\ref{eq:linear_bias_model}  in 
	Figure~\ref{fig:best-fit-biases}  (black dashed line).	We find that across  
	the entire $z$ range the best-fitting values of $b_1$ obtained by modelling the 
	\HI bispectrum are in good agreement with the predictions of Equation~\ref{eq:linear_bias_model}	obtained in Paper I by modelling the 
	\HI power spectrum. Considering the quadratic \HI bias $b_2$ in Figure~\ref{fig:best-fit-biases}  we see that it starts from a value 
	$\approx 3.1$ at $z=6$ and declines rapidly with decreasing $z$, crosses zero 
	at   $z \sim 3$ and then flattens out  at a negative value $\approx -0.36$ at 
	$z<2$. We find that the $z$ dependence of $b_2$ can be very well modelled  using 
	a quartic polynomial  in $z$ containing  only even terms, 
	\begin{equation}
	b_2(z)=b_{20} + b_{22} z^2 + b_{24} z^4\,,
	\label{eq:model_b2}
	\end{equation} 
	with  the fitting parameters  $(b_{20},b_{22},b_{24})=(-0.365, 0.0121, 0.00217)$.

	\begin{figure}
		\centering
		\includegraphics[width=0.47\textwidth,angle=0]{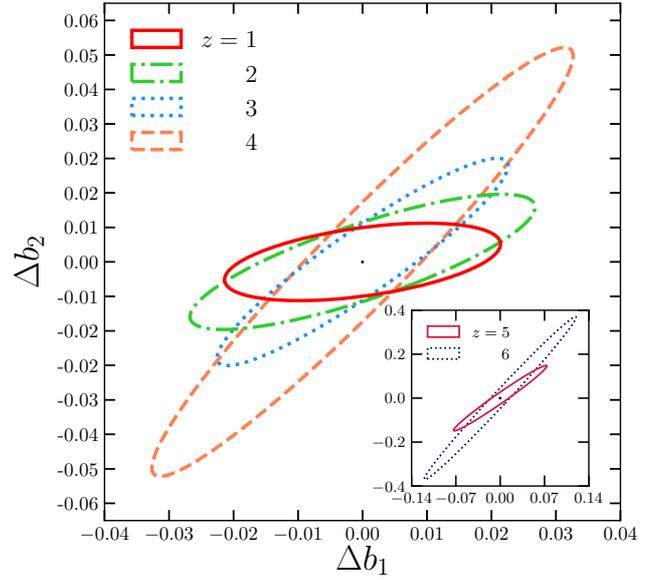}
		\caption{This shows $1-\sigma$ error covariance ellipses obtained 
			from the joint fitting of  the  two bias parameters  $b_1$ and 
			$b_2$ at different redshifts. The error covariance
			ellipses at $z=5$ and $6$ are shown in the inset.}
		\label{fig:err_ell}
	\end{figure}

	Figure~\ref{fig:err_ell} shows $1-\sigma$ error covariance ellipses obtained 
	from the joint fitting of  the  two bias parameters  $b_1$ and $b_2$ at 
	different redshifts. The signal to noise ratio of the estimated bispectrum 
	increases with decreasing redshift. We see that the  errors in  the two 
	bias parameters show the same behaviour {\em i.e.} the errors decrease 
	with decreasing redshift.   The correlation between the errors, as 
	inferred from the tilt of the ellipses, also decreases with 
	decreasing $z$.

	\section{Summary and Discussion}
	\label{sec:discussion}
	We have determined the bispectrum of the post-reionization \HI 21-cm signal 
	using  simulations.	The simulations start from Gaussian initial conditions and 
	the bispectrum emerges from the non-linear gravitational clustering and the 
	non-linear bias of the \HI distribution. We find that the  dimensionless 
	\HI bispectrum 	$\Delta^3_{\HI}(k_1,n,\cth)$ increases monotonically with 
	$k_1$ (Figure~\ref{fig:bispec_k1_z}) and it shows an approximate  power 
	law dependence on the size of the triangle  $\Delta^3_{\HI}\sim k_1^3$ 	(Figure~\ref{fig:k1_dependence})  almost independent of redshift and the 
	shape of the triangle. This scaling is found to be largely restricted to 
	the weakly non-linear regime ($k_1 <1 \mpci $), and the $k_1$ dependence 
	is found to steepen at larger $k_1$ and larger $n$ values where the modes 
	involved  are in the strongly non-linear regime. 
	Considering the $z$ dependence, we find that the amplitude of $\Delta^3_{\HI}(k_1,n,\cth)$ does not evolve significantly with $z$ 
	at $k_1<0.4\mpci$ (Figure~\ref{fig:bispec_k1_z}). We find that  $\Delta^3_{\HI}(k_1,n,\cth)$ shows a weak $z$ evolution at $k_1>0.4\mpci$,
	its amplitude initially decreases with decreasing $z$ reaching a minima 
	at $z\sim3.5$ and then increases at lower $z$ (Figure~\ref{fig:z_variation}). 
	The increase at low $z$ is particularly more pronounced at large 
	$k_1$ and $n$ where the modes involved are strongly non-linear. 
	We see (Figures~\ref{fig:shape_dependence} and \ref{fig:best-fit-bispec}) that
	$\Delta^3_{\HI}(k_1,n,\cth)$ has a ``U" shaped $\cth$ dependence. The ``U" shape 
	is well pronounced when the modes involved are in the weakly non-linear regime. 
	However, this ``U" shape is flattened out at large $k_1$ and $n$ where 
	the modes involved are in the strongly non-linear regime.

	Here  we have  modelled  the \HI bispectrum using  Equation~\ref{eq:HI-bispec-th2} 
	which is based on  second order perturbation theory \citep{scoccimarro00-bispec}.  
	The model expresses the \HI bispectrum in terms of the \HI power spectrum and $b_1$ 
	and $b_2$ which are respectively the linear and quadratic \HI bias parameters. 
	We have used the \HI power spectrum from Paper I, and the model then has 
	only two unknown parameters $b_1$ and $b_2$ which are assumed to evolve with 
	redshift. We find that the model provides a good fit to the \HI bispectrum 
	obtained from the simulations provided we excluded the triangles  where the 
	modes are in the strongly non-linear regime. The linear bias $b_1$ is found 
	to be decreasing nearly linearly with $z$, and the values are in good agreement 
	with the large scale ($k_1 \rightarrow 0$) linear bias estimated in Paper I 
	by directly  modelling  the simulated \HI power spectrum. 
	On the other hand, the best-fitting values of the quadratic bias $b_2$ 
	falls sharply with decreasing $z$, becomes zero at $z\sim 3$ and attains a 
	nearly constant value $b_2 \approx-0.36$ at $z<2$. We have provided polynomial 
	fitting formulae for the $z$ dependence of both $b_1$ and $b_2$. We note that
	$b_2\approx0$ at $z\sim 3$ which implies that the \HI is linearly biased with 
	respect to the underlying dark matter at this redshift.

	The post-reionization \HI 21-cm signal  is a potential tool 
	for precision cosmology. Measurements of the redshift-space 21-cm   
	power spectrum (discussed in Paper II and III) can be used to estimate 
	the  redshift distortion parameter $\beta=f(\Omega_m)/b_1$.  As discussed here,
	measurements of the 21-cm bispectrum can  be used to
	estimate the \HI bias parameters $b_1$ and $b_2$. These  
	can be combined to estimate  the cosmological  growth rate 
	$f(\Omega_m)$ which is not only a  sensitive probe of cosmology  
	but can also be used to distinguish between various theories of gravitation.  
	It is also possible to use the measured 21-cm bispectrum to constrain 
	primordial non-Gaussianity provided one has a  precise model for the  21-cm
	bispectrum expected from Gaussian initial conditions. We plan to investigate 
	these issues in future work. Further, the entire analysis here does not take 
	into account  redshift-space distortion due to peculiar velocities.  
	We plan to incorporate this in future work.


	\bsp	
	\label{lastpage}
\end{document}